\documentstyle[11pt,aaspp4,tighten]{article}

\tightenlines

\def\chs{$\chi^2$ }

\def\gsimeq
{\hbox{\raise0.5ex\hbox{$>\lower1.06ex\hbox{$\kern-1.07em{\sim}$}$}}}
\def\lsimeq
{\hbox{\raise0.5ex\hbox{$<\lower1.06ex\hbox{$\kern-1.07em{\sim}$}$}}}

\begin{document}

\title{Evidence for Relativistic Outflows in Narrow-line Seyfert 1 Galaxies}

\author{Karen M. Leighly}
\affil{Columbia Astrophysics Laboratory, 538 West 120th Street, New
York, NY 10027, USA, leighly@ulisse.phys.columbia.edu}
\author{Richard F. Mushotzky}
\affil{Goddard Space Flight Center, Code 660.2, Greenbelt, MD 20770,
mushotzky@lheavx.gsfc.nasa.gov}
\author{Kirpal Nandra\altaffilmark{1}}
\affil{Goddard Space Flight Center, Code 660.2, Greenbelt, MD 20770,
nandra@lheavx.gsfc.nasa.gov}
\altaffiltext{1}{NAS/NRC Research Associate}
\author{Karl Forster}
\affil{Columbia Astrophysics Laboratory, 538 West 120th Street, New
York, NY 10025, USA, karlfor@mikado.phys.columbia.edu}

\slugcomment{Accepted for publication in {\it The Astrophysical Journal Letters}}


\begin{abstract}

We report the observation of features near 1 keV in the {\it ASCA}
spectra from three ``Narrow Line Seyfert 1'' (NLS1) galaxies.  We
interpret these as oxygen absorption in a highly relativistic outflow.
If interpreted as absorption edges, the implied velocities are
0.2--0.3~c, near the limit predicted by ``line-locking'' radiative
acceleration.  If instead interpreted as broad absorption lines, the
implied velocities are $\sim0.57$~c, interestingly near the velocity
of particles in the last stable orbit around a Kerr black hole,
although a physical interpretation of this is not obvious.  The
features are reminiscent of the UV absorption lines seen in broad
absorption line quasars (BALQSOs), but with larger velocities, and we
note the remarkable similarities in the optical emission line and
broad band properties of NLS1s and low-ionization BALQSOs.

\end{abstract}

\clearpage

\section{Introduction}

Narrow-line Seyfert 1 galaxies (NLS1s) are defined by their optical
line properties (e.g. Goodrich\markcite{11}~1989): ({\it i.}) the
Balmer lines are only slightly broader than the forbidden lines
(H$\beta$~FWHM$<2000\,\rm km/s$); ({\it ii.}) the forbidden line
emission is relatively weak ([O~III]/H$\beta<3$); ({\it iii.}) there
are often strong emission features from Fe~II and high ionization
optical lines.  It has recently been discovered that they also have
distinctive X-ray properties.  {\it ROSAT} PSPC observations found
that the soft X-ray spectra are systematically steeper than
``classical'' Seyfert 1s and that the photon index appears correlated
with the optical line width.  NLS1s also very frequently exhibit rapid
and/or high amplitude X-ray variability (Boller, Brandt
\& Fink\markcite{2}~1996; Forster \& Halpern\markcite{10}~1996 and
references therein).

NLS1 observations with {\it ASCA}, which has better energy resolution
and a larger band pass, find that the steep spectrum in the soft X-ray
band is primarily due to a strong soft excess component with
characteristic blackbody temperature in the range 0.1--0.2~keV and a
relatively weak hard power law.  The hard X-ray power law slope is
either remarkably variable (Leighly et al.\markcite{18}~1996;
Guainazzi et~al.\markcite{12}~1996), or significantly steeper than
found in broad-line Seyfert 1s (Pounds, Done \&
Osborne\markcite{27}~1995; Brandt, Mathur \& Elvis\markcite{7}~1997).
The combination of strong soft excess and steep power law prompted
Pounds, Done \& Osborne\markcite{27}~(1995) to postulate that NLS1s
represent the supermassive black hole analog of Galactic black hole
candidates in the high state.

\section{Data and Analysis}

We considered all the {\it ASCA} data from NLS1s in the archive as of
February 1997, and three proprietary data sets, yielding a sample of
16 objects.  A standard uniform analysis was applied to all data (e.g.
Nandra et al.\markcite{25}~1997; details in Leighly et al.\markcite{19}~1997).

In most cases, a power law model fit the spectra poorly, and an
additional soft excess model was required.  The soft component in
these NLS1s is hot and prominent, and a single black body is not broad
enough to fit the whole energy range from 0.48~keV.  A disk blackbody
model, which is the sum of blackbodies (Makishima et
al.\markcite{22}~1986), is broader.  Therefore, we fit a disk
blackbody model in the energy band $>0.48$~keV. However, our
discussion of absorption features below does not depend critically on
the assumed model of the soft continuum.  Fits over a truncated energy
band $>0.6$~keV with a single blackbody give consistent parameters for
the absorption features, with comparable $\chi^2_\nu$.

In many cases, the power-law plus disk blackbody model gave an
adequate fit.  However, a significant dip remained near 1 keV in the
fit residuals from 1H~0707-495, IRAS~13224-3809, and PG~1404+226
(hereafter H0707, IR1322, and PG1404; Fig.~1; also see
Hayashida\markcite{14} 1996; Otani, Kii \& Miya\markcite{26} 1996;
Comastri, Molendi \& Ulrich\markcite{9} 1997).  Similar features have
been reported in {\it ROSAT} spectra from Akn~564 (Brandt et
al.\markcite{5}~1994) and PG1404 (Ulrich-Demoulin \&
Molendi\markcite{35}~1996).  Addition of an edge to the model improved
the fit by $\Delta\chi^2$/degrees of freedom (d.o.f.)$=33/265$,
$46/307$ and $35/156$, significant at $>99.9\%$ confidence level, for
H0707, IR1322 and PG1404, respectively.  For H0707 and IR1322, an
additional edge reduced the $\chi^2$ by $24$ and $11$, respectively.
The best fit parameters for these two-edge fits, and the single edge
fit for PG1404, are listed in Table~1.

\begin{figure}[t]
\vbox to4.5in{\rule{0pt}{4.5in}}
\includegraphics{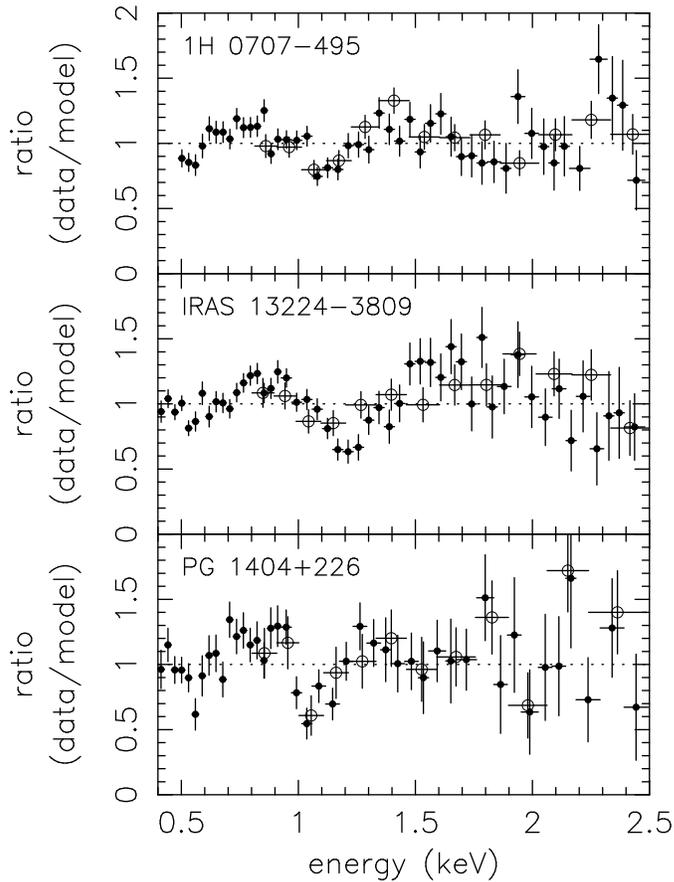}
\caption{Ratio of data to a power-law plus blackbody model, showing
the edge-like feature around 1~keV in these three objects.  Filled and
open symbols denote summed SIS and GIS spectra, respectively.}
\end{figure}

Absorption edges, a signature of highly ionized gas in the
line-of-sight, are commonly found in the X-ray spectra from bright
Seyfert 1 galaxies (e.g. Reynolds\markcite{28}~1997).  Oxygen edges
from O~VII and O~VIII at 0.74 and 0.87~keV respectively are expected
to dominate the absorption profile because of the large abundance and
cross sections.  However, the edges in our objects are found at much
higher energies than could be attributed to oxygen in the AGN rest
frame.  In this energy range, absorption by neon and iron is expected, but
for abundances near solar, it is difficult to produce deep absorption
edges from these elements without also finding strong oxygen edges.  Since
neon differs in atomic number by only two from oxygen, its ionization
state should not be drastically different.  Since both oxygen and neon
are plausibly produced predominately by Type II supernovae, their
relative abundance should not vary much.  Iron species which could
absorb from the L~shell are still present when oxygen is nearly fully
ionized, but the expected dominant ions have higher energy absorption
edges than we find here, near 1.2--1.4 keV.

We can attribute the absorption edges to ionized oxygen if the
absorbing material is being accelerated away from the nucleus, similar
to broad absorption line galaxies (BALQSO; see
Weymann\markcite{37}~1995 and Turnshek\markcite{32}~1995 for recent
reviews).  If we identify the observed edges in H0707 and IR1322 with
O~VII and O~VIII, we infer an ejection velocity of $\sim~-0.2$ to
$-0.3\rm\, c$ (Fig.~2a).  This interpretation is supported by the
consistency in velocity for both ions.  For PG1404, we detect only one
edge, and infer velocity of either $-0.35$ or $-0.20\rm\,c$ depending
on whether it is due to O~VII or O~VIII.

\begin{figure}[t]
\vbox to4.5in{\rule{0pt}{4.5in}}
\includegraphics{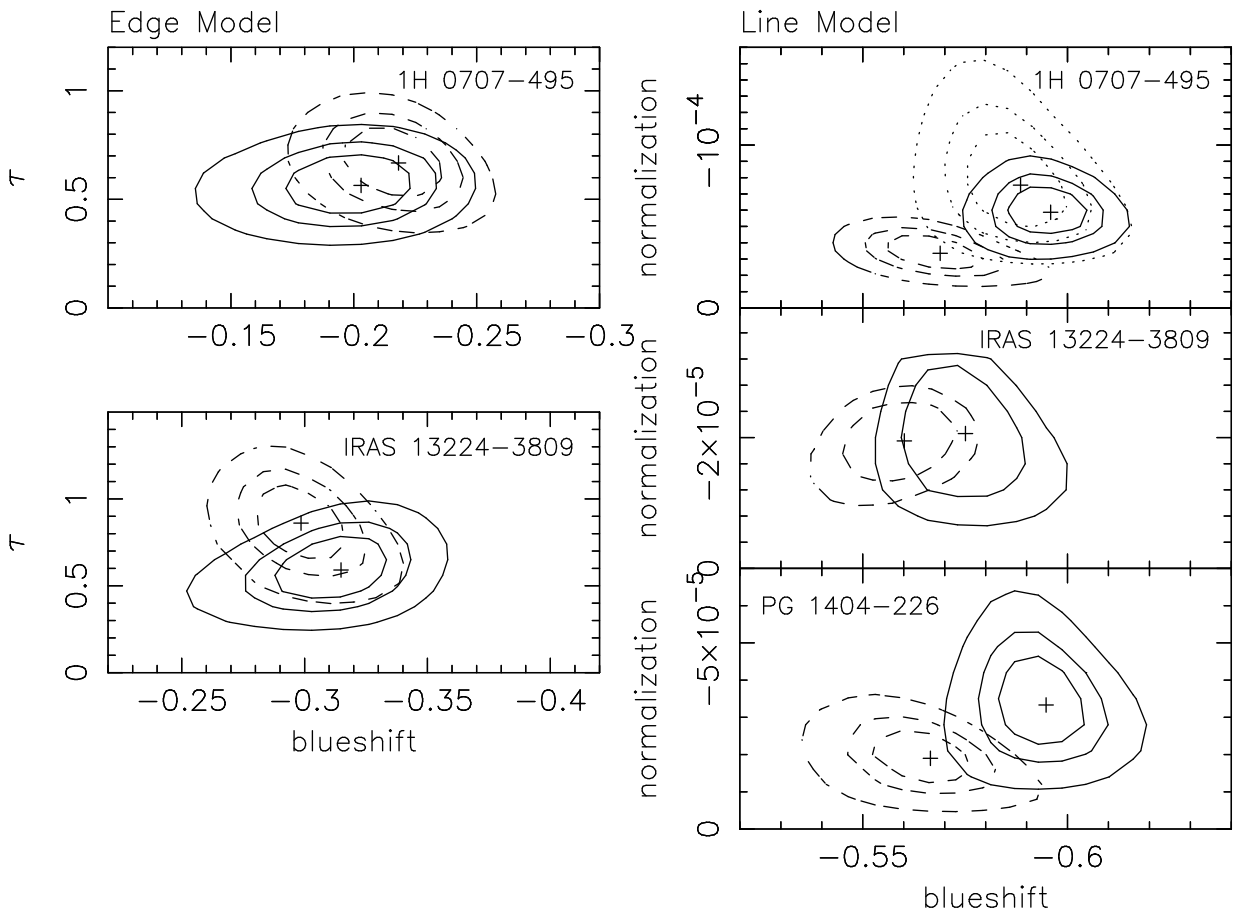}
\caption{When the edges and lines are identified with transitions of oxygen,
the ejection velocities from the nucleus can be found. \chs contours
show 67\%, 90\% and 99\% confidence intervals; for IR1322, only 67\% and
90\% confidence are shown. a.)  Absorption edge fits; b.) Absorption
line fits.  Solid, dashed and dotted lines indicate O~VII, O~VIII, and
N~VI, respectively.}
\end{figure}

Alternatively, the absorption features may be due to resonance line
absorption.  The {\it ASCA} SIS detectors have moderate resolution
($\sim 60$~eV at 1~keV), and since the lines have low equivalent width
$\sim 50\,\rm eV$, they can be detected only if the velocity
dispersion is $\sigma_{vel}
\gsimeq 3000$~km s$^{-1}$. We added ``narrow''
($\sigma=0$) gaussian lines to the power law plus disk blackbody
model, effectively modeling features narrower than the detector
response width.  Addition of an absorption line to the disk blackbody
plus power law model reduced the $\chi^2$ by $30/265$, $25/303$, and
$21/151$~d.o.f., and the second line by $15/263$, $6/301$,
$14/149$~d.o.f. for H0707, IR1322 and PG1404 respectively.  Addition
of a third line for H0707 and IR1322 reduced the $\chi^2$ by $20/261$
and $5/298$~d.o.f. Details of these multiple-line fits are shown in
Table 1.  The final $\chi^2$ are comparable to the edge fit results so
we cannot distinguish these models statistically.  Note that the two
lines present in all spectra are consistent in energy from object to
object, although the differences in cosmological redshift ($z=0.041$,
0.067 \& 0.098 for H0707, IR1322 and PG1404, respectively) imply a
difference of 50~eV in observed energy.  Assuming the absorber is
completely opaque and fitting with the XSPEC {\it notch} model
(equivalent to a very saturated absorption line) gives a velocity
profile FWHM of $\sim 10-14,000\rm\,km/s$, similar to that found in
BALQSOs.  Resonance absorption is expected to be strong in
hydrogen-like and helium-like atoms, and the ratios of the energies
should be the same for every element. Again, we can plausibly identify
the two common lines in all spectra as O~VII and O~VIII at $\sim
0.565$ and 0.651~keV, respectively.  A lower energy line in the H0707
spectrum is plausibly N~VII at 0.498~keV, while a higher energy line
in the IR1322 spectrum has no clear identification.  These line
identifications imply consistent ejection velocities, near
$-0.57\rm\,c$ (Fig.~2b).  Absorption edges should accompany the
resonance absorption lines (Madejski et al.\markcite{20}~1993).
Adding O~VII and O~VIII absorption edges to the model with energies
fixed at the values predicted if $v/c=-0.57$, we find that the optical
depths $\tau$ are consistent with zero, and the $\Delta\chi^2=4.61$
upper limits were 0.27, 0.68 and 0.50 for O~VII and 0.32, 0.13 and
0.26 for O~VIII, for H0707, IR1322 and PG1404 respectively.  In
Section~3 we show that estimations of the column densities derived
from these upper limits are consistent with those from the absorption
line equivalent widths.

As noted above, the measured edge energies seem inconsistent with
those expected from iron-L species. However, iron is a complex ion
potentially producing a complex absorption profile which might fit
these moderate resolution spectra.  We used a power law plus disk
blackbody model, transmitted through an ionized absorber in the AGN
rest frame with iron abundance free (Magdziarz \&
Zdziarski\markcite{21}~1995).  In this model, the electron temperature
$T$ and the ionization parameter $\xi$ are both free parameters, and
we chose two representative values $T_{low}=3\times 10^4$~K and
$T_{high}=3\times 10^5$~K.  These models produced a worse fit overall
compared with the models described above and required rather large
iron overabundances.  For $T=T_{low}$, we obtained iron abundance
lower limits of 15, 29 and 17 times solar with
$\Delta\chi^2$/d.o.f.=11/264, 13/301, \& 6/150, and for $T=T_{high}$,
we obtained iron abundance best fits values of 3.9, 7.3 and 12.8 with
$\Delta\chi^2$/d.o.f.=0/264, 12/301, \& 7/150 for H0707, IR1322 and
PG1404 respectively.  Iron enhancements up to 10 with respect to
oxygen might be expected in some cases, but probably not as large as
20 (Hamann \& Ferland\markcite{13}~1993).  A high iron abundance could
not be responsible for the strong Fe~II emission in NLS1, because the
resulting cooler temperature decreases the number of Fe$^+$ ions while
the optical depth to escaping Fe~II increases (e.g.
Joly\markcite{16}~1993).

In summary, we tried to explain the 1~keV features using three
different models.  Either absorption edges or lines produced a good
fit which could not be distinguished statistically. The iron
overabundance model generally gave a poorer fit.

\section{Discussion}

The outflow velocities of $0.2-0.6\rm\,c$ inferred in these objects
are very large, larger than those observed the UV spectra of BALQSOs.
Such large velocities could be difficult to identify in the UV,
because the prominent C~IV line would be shifted out of the band pass
or into the Ly$\alpha$ forest where it might be hard to distinguish.
Also, very large velocity dispersions with moderate column densities
might result in lines so broad that they blend in with the continuum.

It is beyond the scope of this paper to speculate on the
mechanism required to accelerate material to these very large
velocities, but it is interesting to note that the velocity implied by
the edge fits is close to that seen in the Galactic jet object
SS~433 of $0.26\rm\,c$ and also to the terminal velocity predicted by
``line-locking'' of $0.28\rm\, c$ (Shapiro, Milgrom \&
Rees\markcite{29}~1986).  The larger velocities implied by the
absorption line fits is intriguingly close to the energy of a particle
in circular orbits around a Kerr black hole (e.g.  Shapiro \&
Teukolsky\markcite{30}~1993).  However, it seems difficult to relate
this fact to a physical outflow mechanism.
 
We obtain lower limits on the equivalent hydrogen column.  For the
absorption edge model, assuming cross sections of 2.8 and $0.98\times
10^{-19}\,\rm cm^{-2}$ for O~VII and O~VIII, respectively, and an
oxygen abundance relative to hydrogen of $8.51\times 10^{-4}$, the
O~VII+O~VIII equivalent hydrogen column densities are in the range
$0.4-1.3 \times 10^{22}\rm\,cm^{-2}$.  For the absorption lines,
assumed to be on the linear part of the curve of growth, oscillator
strengths of 0.694 and 0.416 for O~VII and O~VIII were used, giving
O~VII+O~VIII equivalent hydrogen column densities of $1.6-2.1
\times 10^{21}\rm\,cm^{-2}$.  In each case, the column density upper
limits on the absorption edges predicted to accompany these absorption
lines were larger, showing that this model is viable.  While
estimation of the ionization parameter depends on the input continuum
and is beyond the scope of this paper, we note that the O~VIII column
is always larger than the O~VII column, implying a fairly high
ionization parameter.

Rest frame absorption features in the X-ray spectra of broad-line hard
X-ray selected AGN are common (Reynolds\markcite{28}~1997), plausibly
arising in the same material as $z_{em} \approx z_{ab}$ absorption
lines found in the UV (e.g.  Mathur, Elvis \&
Wilkes\markcite{23}~1995).  These `associated' absorption features may
be related to the broad UV absorption lines seen in higher luminosity
objects (e.g. Kolman et al.\markcite{16}~1993).  Evidence suggests
that some aspect of the NLS1 central engine is significantly different
compared with broad-line objects; for example, they may be
characterized by a higher accretion rate relative to Eddington
(Pounds, Done \& Osborne\markcite{27}~1995). The rapid, higher
amplitude and perhaps characteristically nonlinear X-ray variability
may be evidence for strong relativistic effects (Boller et
al.\markcite{1}~1997; Leighly et al.\markcite{19}~1997).  These
results may indicate a higher level of activity relative to the black
hole mass in NLS1s, and strong relativistic outflows might be
expected.

The blue-shifted absorption features discussed here are reminiscent of
those found in the UV spectra of broad absorption line quasars.  A
connection between NLS1s and BALQSOs may be quite reasonable
considering that they have many optical emission line and broad band
continuum properties in common.  Many NLS1s and BALQSOs show strong or
extreme Fe~II emission and weak [O~III] emission.  Objects which have
the strongest Fe~II emission and weakest or no [O~III] emission tend
to be either low ionization BALQSOs or NLS1s (Boroson \&
Meyers~\markcite{4}~1992; Turnshek et al.\markcite{34}~1997; Lawrence
et al.\markcite{17}~1997).  Many NLS1s and low ionization BALQSOs have
red optical continuum spectra, and relatively strong infrared emission
(Boroson \& Meyers~\markcite{4}~1992; Moran, Halpern \&
Helfand\markcite{24}~1996; Turnshek\markcite{33}~1997).  Finally, both
classes are predominantly radio quiet (Stocke et
al.\markcite{31}~1992; Ulvestad, Antonucci, \&
Goodrich\markcite{36}~1995).

An interesting possibility is that the low-ionization BALQSOs and
NLS1s have a common parent population (e.g. Lawrence et
al.\markcite{20}~1997). If so, perhaps objects with intermediate
properties between NLS1s and low-ionization BALQSOs should exist.  It
was recently reported that NLS1 IRAS~13349+2438 has UV broad
absorption lines (Turnshek\markcite{33}~1997).  But while most BALQSOs
are X-ray quiet, this object is a bright soft X-ray source and has the
very steep hard X-ray spectrum and rapid X-ray variability
characteristic of NLS1s (Brinkmann et al.\markcite{8}~1996).

\acknowledgements

KML acknowledges many enlightening discussions with Jules Halpern, and
helpful advice from Tim Kallman.  KML gratefully acknowledges support
through NAG5-3307 ({\it ASCA}). KN thanks the NRC for support.

\newpage

\begin{deluxetable}{llllllllll}
\scriptsize
\tablewidth{0pc}
\tablenum{1}
\tablecaption{Spectral Fitting Results}
\tablehead{
\colhead{Target} & \colhead{$\rm N_H$\tablenotemark{a}} 
& \multicolumn{2}{c}{Power-law} &
\multicolumn{2}{c}{Blackbody} & \multicolumn{3}{c}{Absorption Feature
\tablenotemark{b}} &
\colhead{$\chi^2$}  \\
\cline{3-4} \cline{5-6} \cline{7-9} \\
\colhead {} & \colhead{} & \colhead{Index} &
\colhead{Norm.\tablenotemark{c}} & \colhead{$\rm kT$ (keV)} &
\colhead{Norm.\tablenotemark{d}} & \colhead{Energy (keV)} &
\colhead{Depth} & \colhead{ID\tablenotemark{e}} &
\colhead{} }
\tablecolumns{10}

\startdata

\multicolumn{9}{c}{ABSORPTION EDGE MODEL:} \nl
H0707 &  $0.63^{+0.71}_{-0.05}$ & $2.29^{+0.22}_{-0.21}$ &
$6.0^{+1.9}_{-1.4}$ & $0.20^{+0.02}_{-0.03}$ & $0.26^{+0.64}_{-0.09}$
& $0.91^{+0.03}_{-0.04}$ &  $0.57\pm 0.20$ & O~VII & 318/263 \nl
 & & & & & & $1.09 \pm 0.03$ & $0.67\pm 0.23$ & O~VIII \nl
IR1322 & $0.48^{+0.63}_{-0}$ & $1.97^{+0.26}_{-0.25}$ &
$2.05^{+0.74}_{-0.55}$ & $0.19^{+0.02}_{-0.03}$ &
$0.18^{+0.49}_{-0.07}$ & $1.03 \pm 0.04$ & $0.60^{+0.28}_{-0.25}$
   & O~VII & 339/300 \nl
 & & & & & &$1.18^{+0.04}_{-0.03}$ & $0.87^{+0.31}_{-0.32}$ & O~VIII \nl
PG1404 & $0.55^{+1.67}_{-0.34}$ & $1.67 \pm 0.42$ &
$1.5^{+1.1}_{-0.7}$ & $0.18^{+0.04}_{-0.05}$ & $0.22^{+4.0}_{-0.15}$ &
$1.07 \pm 0.03$ & $1.15^{+0.41}_{-0.44}$  & O~VII or O~VIII & 153/151 \nl
\tableline
\multicolumn{9}{c}{ABSORPTION LINE MODEL:} \nl
H0707 & $1.6 \pm 0.08$ & $2.25\pm 0.18$ & $5.9^{+1.4}_{-1.2}$ & 
$0.15 \pm 0.02$ &
$2.5^{+6.5}_{-1.8}$ &  $0.98\pm 0.03$ & $30^{+20}_{-15}$ & N~VII &  319/261 \nl
& & & & & &  $1.12^{+0.03}_{-0.02}$ & $47^{+18}_{-17}$ & O~VII \nl
& & & &  & & $1.24\pm 0.03$ & $46\pm 20$ & O~VIII \nl
IR1322 & $2.3^{+1.1}_{-0.9}$ & $1.89\pm 0.23$ & 
$1.80^{+0.57}_{-0.46}$ &  $0.13 \pm 0.02$ & 
$5.3^{+22.2}_{-4.1}$  &
$1.09^{+0.05}_{-0.03}$ &  $24^{+18}_{-17}$ & O~VII & 343/298 \nl
& & & & & & $1.23\pm 0.04$ & $48^{+19}_{-24}$  & O~VIII \nl
& & & & & & $1.33^{+0.06}_{-0.05}$ &  $46^{+29}_{-31}$  & ? \nl 
PG1404 & $2.3^{+1.4}_{-1.2}$  & $1.68\pm 0.33$ &
$1.53^{+0.75}_{-0.54}$ &  $0.13^{+0.03}_{-0.02}$ &
$4.8^{+34.0}_{-4.0}$ & $1.12 \pm 0.03$ & $50^{+29}_{-23}$ & O~VII & 150/149  \nl
& & & & & & $1.24 \pm 0.04$ & $51^{+30}_{-27}$ & O~VIII \nl
\enddata
\tablecomments{The quoted uncertainties are 90\% confidence for two
parameters of interest ($\Delta\chi^2$=4.61).  $^a$ Neutral continuum
aborption, in units of $10^{21}\rm\,cm^{-2}$.  The lower limit is the
Galactic absorption column for the edge fits. $^b$ Absorption
feature depths are the optical depth $\tau$ and the line equivalent
width in eV for the edge and line models, respectively. Listed on successive
lines are parameters for models with more than 1 absorption
feature.  $^c$ In units of $10^{-4}\rm\,photons\,cm^{-2}\,s^{-1}$ at
1~keV. $^d$ Multiple black body model (Makishima et al. 1986) $\times
10^{3} (R_{IN}/D_{10})^2*\cos(\theta)$, where $R_{IN}$ is the inner
disk radius in kilometers, $D$ is the distance to the source in units
of 10 kiloparsecs, and $\theta$ is the disk inclination. $^e$
Tentative edge/line identification, implying velocities in Fig.~2.}
\end{deluxetable}
\clearpage

\end{document}